# Raman scattering of graphene based systems in high magnetic fields

Clement Faugeras,* Milan Orlita, and Marek Potemski

LNCMI (CNRS, UGA, UPS, INSA, EMFL), BP 166, 38042 Grenoble Cedex 9, France
E-mail: clement.faugeras@lncmi.cnrs.fr

## Abstract

We review the different results obtained in the last decade in the field of Raman scattering of graphene based systems, with an applied magnetic field. Electronic properties of graphene based systems with an applied magnetic field are first described. The phonon response in magnetic field, the magneto-phono resonance, is then introduced and described for graphene mono- and multi-layers, as well as for bulk graphite. Electronic Raman scattering are then discussed in the context of Landau level spectroscopy, of electron phonon interaction and of electron-electron interaction.

## Introduction

Raman scattering is a very important and popular characterization technique in the field of research of carbon-based material,[1] including graphene mono- and multi-layers, bulk graphite, carbon nanotubes, fullerenes, and also amorphous carbon.[2,3] The main contribution to the Raman scattering spectrum of such materials is due to lattice vibrations, phonons, which are characteristic of the hybridization of the carbon atoms, of the dimensionality of the system, as well as on the polytype. This technique offers a non-destructive and, in a first approximation, non-invasive probe to study and characterize this whole family of materials.

Optical spectroscopy, combined with high magnetic fields environments, is a very efficient tool to explore the optical and electronic properties of solids. Landau quantization in electronic systems, together with the evolution with magnetic fields of the resulting Landau levels, are characteristic of the electronic band structure. Landau level spectroscopy, that traces for instance the evolution of Landau levels by scanning tunneling spectroscopy, or of inter Landau level excitations by optical means, gives direct information about the band structure, about scattering rates and mechanisms, and about doping levels. Applying a magnetic field provides a tool to tune the energy of Landau levels and of the associated inter Landau level electronic excitations. Hence, the energy of electronic excitations can be monitored externally, and in particular, it can be tuned to the energy of other excitations of the system like phonons, in search for interaction effects.

In this article, we start by introducing the basics of electronic and phonon properties of graphene based systems in order to discuss magneto-phonon resonances (MPR) phenomena. The case of graphene mono- and multi-layers together with bulk graphite will be discussed. MPR allow studying the electron-phonon interaction and also performing the Landau level spectroscopy of electronic excitations at the optical phonon energy. We then focus on the contribution of electronic excitations to the Raman scattering spectrum, that is the direct observation of inter Landau level electronic excitations and the study of their evolution with increasing magnetic fields. We finish by presenting prospects concerning Raman scattering of graphene based systems in magnetic fields.

## Magneto-phonon resonance

Graphene is the building block of all sp$^2$ carbon allotropes. A large part of the richness of the Raman scattering spectrum of these materials comes from the gapless or semi-metallic nature of graphene, and from the existence of two inequivalent K and K' valleys. These two valleys allow second order scattering processes, the so-called double resonances,[4] which connect the excitations laser energy to the phonon and the electronic band structures. Some Raman scattering features are indeed combinations of electronic and phonon or of two phonons excitations. The Raman scattering spectrum can trace the evolution of the electronic band structure when changing the number of graphene layers[5,6] or of their stacking,[7,8] to study the phonon dispersion including phonons away from the Γ point, and to get a precise picture of the type and density of scattering centers.[9-12]

The doubly degenerate optical phonon at the Γ point of the phonon Brillouin zone with the highest energy, the E$_{2g}$ phonon (see Fig. 1a) gives rise to the G band feature in the Raman scattering spectrum, close to 1580 cm−1. The energy of this phonon is sensitive to temperature,[13] to strain,[14,15] and, unique to this family of materials, to the electronic excitation spectrum.[16-19] This last point is rather unusual but in graphene, lattice vibrations are effectively screened by electronic excitations at the particular Γ and K points of the phonon Brillouin zone. This abrupt change of the phonon energy and dispersion due to variations of the screening are called Kohn anomalies.[20] They can be evidenced by measuring directly the phonon dispersion,[21] or by tracing the evolution of the G band when modifying the electronic excitation spectrum, for instance by tuning, at $B$ = 0, the position of the Fermi level and sequentially quenching low energy electronic excitations due to the Pauli principle.[17,18]

Another method to modify the electronic excitation spectrum is by applying a magnetic field perpendicular to the graphene planes. For clean enough electronic systems, Landau quantization occurs and the density of states, which grows linearly with the energy in the absence magnetic field, transforms into a series of discrete and highly degenerate Landau levels. The details of the Landau level energy ladder are unique for each electronic band structure and the knowledge of this energy ladder brings a similar knowledge on the zero field electronic dispersion. For a graphene monolayer, the Landau level spectrum is expressed by $E_n = \pm v\sqrt{2e\hbar B}\sqrt{|n|}$, where the ± describes the conduction and valence bands, v ~1×10$^6$ m.s$^{-1}$ is the band velocity, B the magnetic field, and $n$ = 0, 1±, 2±, ... is the Landau level index. Graphene bilayers also show a simple Landau level spectrum expressed by $E_n = \pm \hbar\omega_c\sqrt{n(n+1)}$ (for E$_n$ ≪ $\gamma_1$) where n = 0, 1±, 2±,... and $\omega_c = 2(\frac{v}{l_B})^2/\gamma_1$ where the magnetic length $l_B = \sqrt{\hbar/(eB)}$ and $\gamma_1$ ~400 meV is the interlayer hoping integral. For

$n > 2$, this energy ladder has the form $E_n \cong \pm\hbar\omega_c(n+\frac{1}{2})$ typical of electrons in parabolic bands. Graphene multilayers with more than 2 layers have electronic spectra which can be in the first approach viewed as a monolayer and of effective graphene bilayers band structures,[23,24] with effective $\gamma_1*$ coupling parameters.[25] Bulk graphite is a 3D system with an electronic dispersion also along $k_z$. The optical properties are dominated by the *H−K−H* corners of its hexagonal Brillouin zone. The in-plane ($k_x$, $k_y$) band structure at the *H* point is very similar to that of a graphene monolayer but with an additional double degeneracy, and the *K* point in-plane dispersion is an effective graphene bilayer with $\gamma_1* = 2\gamma_1$. Application of a magnetic field along the c axis of bulk graphite creates Landau bands, which are highly degenerated but are not discrete as in the case of 2D and quasi-2D systems. They disperse along the direction of the applied magnetic field and have an energy extend at a fixed magnetic field perpendicular to the plane of the layers, imposed by the energy dispersion along $k_z$.

Within these Landau level ladders, some particular inter Landau level excitation are Raman active, some are infrared active and some couple to optical phonons. The Landau level index is related to the angular momentum and inter Landau level excitations $L_{n,n'}$, where n and n' are the indices of the two Landau levels involved in the excitation, carry an angular momentum $m_z = |n'|-|n|$. The optical selection rules for light absorption (infrared absorption) are Δ|n| = ±1, corresponding to the two possible circular polarization of light σ±.[26] The Raman scattering selection rules, presented in Fig. 2, are different and they take into account both incoming and outgoing photon polarization.[27-29] If these polarizations are identical, then the total angular momentum transfer is zero and Δ|n| = 0 inter-Landau level excitation are active. These excitations are the more intense and represent the main contribution to the electronic Raman scattering spectrum. If the circular polarizations of the incoming and outgoing photons are different, then the angular momentum transfer is $m_z$ = Δ|n| = ±2. These excitations are predicted to weakly contribute to the Raman scattering spectrum, but because trigonal warping reduces the angular momentum space to $m_z$ = (0, 1, 2),[27] they give some Raman activity to Δ|n| = ±1 excitations. These latter excitations have the same energy as optical excitations and, because they share the same symmetry as the Γ-point optical phonon, they strongly couple to the G band and are the relevant excitations for MPR.

Magneto-phonon resonance describes a class of phenomena, resulting from the coincidence of the energy spacing between two Landau levels of an electronic system subjected to a magnetic field, and the energy of an optical phonon. MPR is seen in the magneto-transport properties of bulk semiconductors and quantum wells as magneto-oscillations of the electron scattering rate.[30,31] Magneto-polarons, the mixed electron-phonon quasi-particle that form when the inter Landau level energy spacing coincides with the phonon energy, have been chased for in magneto-infrared spectroscopy.[32] In this context, the case of graphene is of particular interest because

i) the inter Landau level energy spacing follows a √n scaling which significantly enlarges the coincidence possibilities, and ii) because of the Kohn anomaly that makes the energy of the optical phonon sensitive to changes of the electronic excitation spectrum.

## Neutral graphene

Under magnetic field, it has been anticipated by two groups independently[22,33] that a magneto-phonon resonance effect should occur in graphene involving the Raman active $E_{2g}$ phonon and $\Delta|n| = \pm1$ inter Landau level excitations, see Fig 1c and d.

MPR are encoded within the phonon polarization operator $\Pi(\omega)$. The coupled electron-phonon response can be expressed by finding the poles $\tilde{\epsilon} = \epsilon - i\Gamma$, where $\epsilon$ is the phonon energy and $\Gamma$ a broadening parameter, of the phonon Green function by solving the following equation:

$$\tilde{\varepsilon}^2 - \varepsilon_0^2 = 2\varepsilon_0 \lambda E_1^2 \sum_{k=0}^{\infty} \left\{ \frac{T_k}{(\tilde{\varepsilon} + i\delta)^2 - T_k^2} + \frac{1}{T_k} \right\}$$

where $\varepsilon_0$ is the phonon energy of the neutral system at $B$ = 0 T, $\delta$ accounts for the broadening characteristic for electronic excitations of energy $T_k = L_{-k,k+1} = E_k + E_{k+1}$, and $E_1 = v\sqrt{2e\hbar B}$. The relevant experimental parameters that enter the phonon polarization operator are the dimensionless electron-phonon coupling constant $\lambda$ and the band velocity $v$. The conditions for the proper observation of MPR are low doping level,[34] spatial homogeneity of the doping profile and low disorder to ensure a reduced broadening of electronic excitations (low $\delta$).

MPR in graphene have been first observed in quasi-neutral graphene systems, namely in multilayer epitaxial graphene on the carbon face of 4H-SiC[35] and decoupled graphene flakes on the surface of bulk graphite.[36-40] An example is presented in Fig. 3, which shows the low temperature evolution of the G band as a function of the magnetic field for graphene on graphite. For neutral graphene systems under magnetic field, the Fermi energy is pinned to the $n$ = 0 Landau level and all interband inter Landau level electronic excitations are Raman active when they are tuned to the phonon energy, see Fig. 1c. This situation brings the richest picture of MPR. Tracing the evolution of the G band as a function of the magnetic field, a series of avoided crossings is observed when the different $L_{n',n}$ excitations are tuned in resonance with the phonon. The most developed avoided crossing happens close to $B$ = 28 T in graphene on graphite and is due to the fundamental $L_{0(-1),1(0)}$ excitation. From this figure, it can be seen that not only $\Delta|n| = \pm1$ excitations couple to the phonon with a strength (splitting energy at resonance) that grows with the magnetic field, but also $\Delta|n| = \pm2$ as seen at $B \sim 14$ T with $L_{0,2}$ crossing the phonon feature, and even more surprising, $\Delta|n| = 0$ excitations. This coupling with $\Delta|n| = 0$ excitations, forbidden for symmetry reasons, is still puzzling and has also been observed in graphene encapsulated in hexagonal Boron Nitride (hBN).[41] When all electronic excitations have energies different than the phonon energy, the phonon line width is minimal and is only imposed by scattering mechanisms other than the electron-phonon interaction and is hence representative of the electronic quality of the material. For graphene on graphite, which is characterized by a very high electron mobility,[42] phonon line width as low as 2.5 cm$^{-1}$ have been reported.[39] The analysis of the MPR effect in graphene brought values for $\lambda = \sim 4 \times 10^{-3}$, in agreement with determinations of this parameter by Raman scattering at $B = 0$[17] or by DFT calculations[43] and electronic broadening parameters close to 90 cm$^{-1}$.

Doped graphene

The case of doped graphene contains new physics compared to neutral graphene because of i) the quenching of some interband excitations due to Pauli blocking, and of ii) intra-band excitations (cyclotron resonance-like) that only exist for doped graphene. The study of these two effects on the MPR requires polarization resolved Raman scattering measurements as a finite doping creates an asymmetry between conduction and valence bands. When graphene is deposited on a standard SiO2/Si substrate, it is often doped and suffers from the interaction with the substrate which induces electron-hole puddles,[44,45] doping spatial inhomogeneities on a length scale of few tens of nanometers, which can completely erase the MPR due to Pauli blocking. Signatures of the MPR in doped graphene have been obtained using polarization resolved Raman scattering techniques and a graphene monolayer deposited on $SiO_2$/Si and naturally p doped at a level close to $2 \times 10^{12}$ cm$^{-2}$. This MPR is due to a single electronic excitation at high magnetic field, the $L_{-1,0}$, because all other excitations including $L_{0,1}$ are Pauli blocked when tuned to the phonon energy.[34]

To go further in the characterization of this effect as a function of the doping, one needs a mean to change the doping of a graphene flake. This was achieved either by thermal annealing[47] or by using structures with an electrostatic gate.[46,48,49] Tuning the position of the Fermi energy changes the occupation of Landau levels and hence, the strength of the electronic excitations involving the Landau level at the Fermi energy. The splitting energy at the resonant magnetic field directly depends on this occupation factor.[22] When setting the magnetic field so that there is a resonance between one $L_{n',n}$ and the phonon, the interacting state of the phonon can be tuned from non-interacting when the electronic excitation is Pauli quenched, to strongly-interacting when the final Landau level of the electronic excitation is completely depleted and the excitation has a maximum strength. On the top of the possibility to externally tune the phonon interaction state, such experiments allowed to highlight the effect of intraband excitations, the cyclotron resonance modes, on the phonon energy. This effect is shown in Fig. 4 which shows for selected values of the magnetic field, the evolution of the phonon energy as a function of the filling factor in a range where all electronic excitations have an energy much bigger than the one of the phonon, except for the $L_{0,1}$ excitation which is not active, and for the cyclotron resonance mode. The effect of this latter excitation appears as a 8 − 10 cm$^{-1}$ energy difference of the phonon energy when measured for opposite values of the filling factors, for instance at $v$ = ±6. This energy difference grows with increasing magnetic fields and is due to the cyclotron resonance mode, which is only active in one polarization configuration for a given value of the filling factor. The possible electronic excitations in the two crossed circular polarization configurations are presented in the right part of Fig. 4, showing how polarization resolved experiments can distinguish between the two types of doping. The complete optical phonon polarization operator, including both interband and intraband electronic excitations and polarization of the incoming and outgoing photons was developed in Ref. 46. Uniaxial strain, very likely to affect realistic samples at low temperatures, has also been invoked to explain details of the evolution of the MPR.[47,50,51]

## Graphene multilayers - bulk graphite

The electronic band structure of graphene multi-layers grows in complexity with the number of layers. The electron-phonon interaction is not modified when increasing the number of graphene layers ($\lambda$ is constant for all sp$^2$ carbon allotrope) and the phonon polarization operator grows in complexity, changing form to take into account the particular electronic excitation spectrum and the overlap of electronic wave function with the different phonons in each specific multilayer structure.[52] MPR can be used as a tool to perform the Landau level spectroscopy of $\Delta|n|$ = ±1 electronic excitations, at the G band energy. The first step of this analysis, is a frequency analysis of the phonon oscillations based on Fourier transform of the scattered intensity at the phonon wave number at $B$ = 0 T. The number of observed frequencies and their absolute values bring the number of excitation spectra composing the signal and hence, the number of layers. This methodology has been developed to determine the number of layers in unknown samples,[52-54] or to investigate velocity renormalization effects in graphene encapsulated in hBN.[41] We present in Fig. 5 a comparison of the MPR on the G band feature of a) a monolayer graphene encapsulated in hBN, b) a bilayer graphene encapsulated in hBN, and c) a quadrilayer graphene deposited on SiO$_2$/Si. The crossing points indicated by white arrows are characteristic of the electron excitation spectrum which can then be precisely determined, allowing for the identification of the sample thickness, or for the determination of band structure parameters. Each graphene multi-layer presents a unique MPR pattern.

The 3D polytype of sp$^2$ carbon, bulk graphite, also shows a MPR involving the E$_{2g}$ phonon and inter Landau band excitations arising mainly from the K point. The in-plane dispersion at the $K$ point of graphite is similar to the one of bilayer graphene, but with an effective interlayer hopping integral $\gamma_1{}^*$ twice enhanced with respect to $\gamma_1$ in a truly bilayer graphene.[55,56] Because of the dispersion along $kz$, Landau bands are formed in contrast to Landau levels in 2D and quasi-2D systems. As a results, inter-Landau band excitations, even though peaked at the transition energy exactly at the $K$ point where the joined density of state diverges, are spread over a broad range of energy (see Fig. 6b). The electron-phonon interaction is then never completely resonant as the electronic mode is not a discrete state. The phonon oscillations, even though less developed than in monolayer graphene, can never the less be observed as it is shown in Fig 5d). The comparison of the MPR measured in the two crossed-circular polarizations in bulk graphite allowed determining the electron-hole asymmetry in this material.[29,57]

## Electronic Raman scattering

To our knowledge, the first report of a purely electronic contribution to the Raman scattering spectrum of a polytype of $sp^2$ carbon was in bulk graphite.[58] Because at that time, the Raman scattering selection rules[27,28,59] were not completely established and because of the now obvious similarities between K-point electrons in bulk graphite and the excitation spectrum of graphene bilayers, this key observation is nowadays interpreted alternatively. It never the less stimulated many studies of electronic excitations in graphene based systems. [29,37,40,57,60-63]

The Raman scattering selection rules for electronic excitations in graphene monolayers[27,59] are described in Fig. 2, and are the same for multilayers[28] and bulk graphite,[29] within the corresponding Landau levels energy spectrum.[23,24] Electronic excitations are best observed in the high quality graphene on graphite system,[42,64] which can be used as a platform to test Raman scattering selection rules. It appears that in this graphene system which shows the richest excitation spectrum, the co-circular polarization configuration selects $\Delta|n| = 0$ electronic excitations and all $\Delta|n| = \pm 1$ excitations except the ones involving the $n = 0$ Landau level.[37,39] The cross circular polarization configuration selects the $E_{2g}$ phonon, the associated MPR including signatures of $\Delta|n| = \pm 2$ (see for instance Fig. 3 at $B = 14$ T), and $L_{0(-1),1(0)}$ excitations.[39] All other graphene systems that showed electronic Raman scattering up to now (suspended graphene and graphene on hBN) only showed $\Delta|n| = 0$ excitations in the co-circular polarization configuration and the MPR in the cross-circular polarization configuration.

The Landau level spectroscopy of graphene can be performed with infrared transmission spectroscopy on macroscopic specimens[65,66] or on specially designed exfoliated structures,[67] but as Raman spectroscopy is performed in the visible range of energy it naturally gives access to micrometer spatial resolution and to the possibility of investigating exfoliated structures. Graphene on bulk graphite[37,40] or sophisticated graphene structures, suspended[60,63] or encapsulated in hBN,[62] have been investigated using Raman scattering techniques and have indeed shown electronic excitations. Fig. 6a), representative polarization resolved Raman scattering spectra of graphene on graphite measured at low values magnetic fields are presented. The $E_{2g}$ is only observed in the crossed circular polarization, and this phonon is Raman active tanks to trigonal warping. In the co-circular polarization configuration, a series of discrete lines with a lorentzian line shape and an energy growing with the magnetic field can be observed. Their full width at half maximum is ~ 40 cm$^{-1}$. On the basis of the Landau level ladder of graphene, and of the Raman scattering selection rules for graphene,[27] it is possible to identify these excitations to the $L_{-n,n}$ series, expected to be the most pronounced excitations in the electronic response of graphene. In Fig. 7, the Landau level spectroscopy of suspended mono-, bi-, tri-, quadri- and penta-layer graphene is presented in the form of a false color map of the scattered intensity as a function of the magnetic field (left) and the extracted peak position is presented on the left side of the figure with calculated dispersions.[60] The use of suspended structures is crucial as it ensures a very low level of doping[68] and a high electronic quality because of the lack of interaction with a substrate. These results clearly demonstrate the potential of Raman scattering to perform the Landau level spectroscopy of micrometer size flakes (suspended in this case), and to determine the relevant parameters characterizing the electronic bands. In the case of multi layers, the field evolution of the different excitations could be well reproduced with a single magnetic field independent $\gamma_1$ parameter, and a number-of-layer dependent band velocity, indicated in Fig. 7, that could be related to screening effects when increasing the number of layers. The suspended monolayer showed surprising effects: a magnetic field and Landau level index dependent band velocity.

## Landau level spectroscopy and electron-electron interaction

Graphene is expected to be a strongly interacting material as the fine structure constant $\alpha$ of electrodynamics, the ratio of Coulomb and kinetic energies, is re-scaled to $\alpha(c/v) \sim 2$ in graphene. Magneto-optical response of semiconductor has most often been interpreted in the frame of Kohn's theorem,[69] which holds for systems with translational symmetry, with a parabolic electronic dispersion, and it only concerns the intra band response, the cyclotron resonance mode. Graphene clearly does not fulfil all these requirements, the most obvious one being the lack of parabolic dispersion. It is then expected that electron-electron interaction modify the magneto-optical excitation spectrum of graphene, also the long wavelength limit. Unfortunately, most of infrared magneto-transmission spectra have been obtained on epitaxial graphene on SiC[65,66] which shows an inter Landau level excitation spectrum in nearly perfect agreement with the single particle picture. Similar experiment performed on graphene deposited on $SiO_2$/Si showed some more exotic behavior [67,70] with a Landau level index dependent band velocity. It is now understood that electron-electron interaction in graphene can be efficiently screened, which is the case in multilayer epitaxial graphene due to the highly doped buffer layer, and in graphene on graphite because of the free carriers in bulk graphite.

In graphene, electron-electron interactions do affect electronic properties.[71-73] Inter Landau level excitations, magneto-excitons, are predicted to have an in-plane dispersion that has been evaluated within various approximations.[74-76] It is however not possible today to access this dispersion experimentally as it has been done in conventional two dimensional electron gas in III-V semiconductor quantum wells.[77,78]

Raman scattering is a probe of long wavelength excitations, at k = 0, and because of its linear dispersion, the long wavelength excitations should also be affected by electron-electron interactions. Indeed, it has been predicted for neutral graphene[79,80] and observed experimentally in magneto-transport of suspended graphene[81] as well as in ARPES spectra of graphene on different substrates,[82,83] that the electronic dispersion is affected by electron-electron interaction. Interactions renormalize the band velocity which acquires an energy dependence. The dispersion changes from linear to logarithmic.[81]

Effects of electron-electron interaction have been searched for in the magnetic field evolution of the inter Landau level excitations for graphene monolayer specimens in different dielectric environments, suspended graphene (G-S), graphene in hBN (G-

BN) and graphene on graphite (G-Gr). The two main excitations $L_{-1,1}$ and $L_{-2,2}$ could be observed in these three structures and their energy has been expressed in the form of the band velocity. The evolution of this effective band velocity as a function of the magnetic field is presented in Fig.8, which clearly shows the three main points, incompatible with the single particle picture: i) strong variation of $v$ for these graphene specimens in different environments, ii) a magnetic field dependence of this effective velocity, together with iii) a Landau level index dependence. To explain these results, first order perturbation theory was used to express the expected evolution for the band velocity as a function of the magnetic field: [62,84]

$$v_n = \frac{\omega_{-n,n}\ell_B}{\sqrt{8n}} = v_0 + \frac{\alpha c}{4\varepsilon}\left(ln\left(\frac{W\ell_B}{\hbar v_0}\right) - ln\left(\frac{\ell_{B_0}}{\ell_B}\right)\right) + \frac{\alpha c}{4\varepsilon}C_n \quad (1)$$

where $v_0$ the bare velocity, $v_n$ the velocity associated with the excitation $L_{-n,n}$ with energy $\hbar\omega_{-n,n}$, W is the high energy cut-off, $\epsilon$ is the dielectric constant of the surrounding medium, c is the speed of light, $B_0$ is an arbitrary reference magnetic field, and the $C_n$ numerical coefficients include the vertex corrections (excitonic effects) together with some contributions to the self-energies. The $C_n$ coefficients reflect the excitonic effect and explain the puzzling ordering of the velocities associated with n = 1 and n = 2. From these results the dielectric constant of graphene in a medium $\epsilon_{env}$ can be expressed as $\epsilon_G$ = $\epsilon_{env}$ + 3 where the factor 3 describes a self-screening from the graphene carriers.

## Bulk graphite

Electronic Raman scattering has also been investigated in bulk graphite,[29,58] showing a very rich and still only partially understood excitation spectrum. Electronic excitations within the Landau bands of graphite have a particular, asymmetric line shape, representative of the joint density of states,[55] and involve electronic states along $k_z$ with varying in-plane electronic dispersions. This material may serve as a textbook example of Raman scattering selection rules: $\Delta|n|$ = 0 excitations are the dominant feature of the spectrum, $\Delta|n|$ = ±2 electronic excitations can also be observed in the crossed circular polarization and they do not couple to the G band, $\Delta|n|$ = ±1 are hardly visible, but they give rise to the MPR. The study of electronic Raman scattering in bulk graphite brought a detailed picture of the band structure (tight binding parameters entering the SWM model[85,86]) including the electron-hole asymmetry which is significant in graphite, and of the MPR.[29,57] These excitations are presented in Fig.6b) at $B$ = 0, 12 and 20 T, together with calculations of the joint density of states.[29] One can note on this figure the scattered amplitude at $B$ = 0 vanishes when a magnetic field is applied. This is a signature of the featureless electronic response of electrons[87] at the K point of bulk graphite, that transforms into inter Landau bands excitations when a magnetic field is applied. Because of the spread of the weight of electronic excitations over 200 − 300 cm$^{-1}$ and because of their asymmetric line shape, oscillations of the phonon line width also reflect this particular line shape.[29] The analysis of the line shape of these inter Landau band excitations in graphite when changing the temperature showed that this line shape reflects the joint density of states, but it was claimed that it is also imposed by electron-electron interactions. The Tomonaga-Luttinger theory, typical of 1D electronic system, was used to describe the observed evolutions.[88]

## MPR beyond Γ-point phonons

In the first part of this article, we have been discussing experiments tracing the evolution of the G band as a function of the magnetic field. When it was realized that inter Landau level excitations could be observed directly and that their evolution with increasing magnetic field could be traced, our understanding of MPR in graphene based systems evolved significantly. It is possible to search for effects related to the electron-phonon interaction directly on the electronic excitation spectrum and to observe effects related to phonons or to multi-particle complexes which are not Raman active themselves,[89] but that can be tuned in resonance with Raman active modes and interact with them.

The results of such experiment, performed on graphene on bulk graphite are presented in Fig. 9. In this figure, blue dots indicate MPR related to Γ-point phonons, red dots MPR related to K point phonons. First, effects of electron-phonon interaction are observed for K point phonons, implying that the electronic excitation also carries an opposite momentum of the order of K to fulfil momentum conservation law and is hence an inter valley excitation. The main effect illustrated in Fig. 9 is the observation of pronounced discontinuities when $\Delta|n| = 0$ electronic excitations are tuned to specific values of energy corresponding to $2\hbar\omega_\Gamma$ and to $2\hbar\omega_K$. This effect can be seen as a strong variation of the excited state life time (n-th Landau level) at the resonant magnetic field. Indeed, when $L_{0,n} = \hbar\omega_{phonon}$, a new scattering channel involving phonons is becomes active, connecting the n-th Landau level to the n = 0 and the lifetime of all electronic excitations involving the n-th Landau, $L_{-n,n}$ $L_{-n+1,n}$ and $L_{-n-1,n}$, level is strongly decreased. This leads at first order to a broadening of the $L_{-n,n}$ line at the resonant magnetic field, together with a simultaneous broadening of $L_{-n+1,n}$ and of $L_{-n-1,n}$.

This effect is in fact more complex as triple anti-crossings are observed at the resonant magnetic fields, indicating a real hybridization of the different modes and not only the opening of an additional scattering channel. At these specific values of the magnetic field, we have the following condition $2\hbar\omega_{Ph} = L_{-n,n} = L_{0,n} + \hbar\omega_{Ph}$, where $\hbar\omega_{Ph}$ is the phonon energy. These three excitations interact through the electron-phonon interaction and produce this triple avoided crossing pattern. At the same values of magnetic field, all observable electronic excitations involving the same Landau level show and avoided crossing behavior. This is well visible at $B \sim 9$ and 18 T for $K$-point phonons, and at $B \sim 14$ and 18 T for Γ-point phonons. When the above mentioned triple resonant condition is fulfilled, we also have $L_{-n,n+1} = L_{0,n+1} + \hbar\omega_{Ph}$ and $L_{-n-1,n} = L_{-n-1,0} + \hbar\omega_{Ph}$. Three different electronic excitations are hence affected by interaction effects at the same value of the magnetic field, and this holds for both Γ and $K$ point phonons. Multi-particle processes are involved in the relaxation mechanisms of hot carriers in graphene, which is of particular importance when considering applications in optoelectronics.

## Prospects

The optical study of MPR with Raman scattering techniques is now well advanced, and in some sense is now part of characterization techniques allowing us to extract physical parameters such as phonon broadenings or to characterize Landau level width.[41] It has been realized recently that strain fields in graphene are equivalent to a pseudo-magnetic fields[90] and can profoundly modify the electronic excitation spectrum[91] by inducing Landau quantization without externally applied magnetic field. MPR can be used as a probe of strain in a well-defined regime where MPR is active, and combined with the micrometer scale spatial resolution achievable with visible optics, MPR can bring a precise spatial picture of strain distribution as was recently proposed in Ref. [92,93]. Interesting and still unexplored yet is the regime of very high magnetic fields, for which the cyclotron resonance mode, the intra band excitation in a doped graphene sample, is tuned to the G band energy. Mainly because it occurs at magnetic fields above $B$ = 100 T, this coupling is expected to give the most pronounced anti-crossing and mode-mixing, but such high fields are today only achieved with pulsed magnetic field installation, and are not compatible with typical integration times of Raman scattering experiments.

MPR of low energy phonons, such as the rigid layer shear and breath modes of graphene multi-layers.[97] The shear mode shares the same symmetry as the G band and should also strongly couple to electronic excitations. To our knowledge, this effect has not been investigated yet and still represents an experimental challenge as it requires low energy magneto-Raman scattering with a micrometer scale spatial resolution.

Electronic Raman scattering in graphene is a growing field in graphene research, which basis have been set in the last years. The featureless electronic response of graphene at B = 0 has been detected recently.[87,95] It is now possible to investigate subtle effects such as dispersions of elementary electronic excitations in magnetic fields,[77] or to investigate specific regimes such as the composite fermion regime at temperatures below T = 1 K. Many body effects have also been invoked to explain the transport gaps observed in graphene multilayers.[53,96] The study of these low energy gaps with Raman scattering techniques may bring new information concerning electron-electron interaction in these materials and the investigation of Landau quantization at low energies can bring details about the structure of the gap and the electronic dispersion in the close vicinity of the band extrema. A band gap with low energy has been recently observed in heterostructures of graphene and hexagonal boron nitride.[97,98] In the same spirit, the study of these energy gaps with low energy magneto-Raman scattering techniques in these simple graphene/hBN heterostructures will be very interesting.

The field of 2D materials has grown since the discovery of graphene and now includes monolayers of transition metal dichalcogenides (TMD), some of them being 2D semiconductors. The band structure of these materials is non-conventional with two inequivalent valleys and strong spin-orbit interaction. Inter band inter Landau level excitations have recently been observed in the low temperature reflectivity spectra of WSe$_2$ in magnetic field.[99] The electronic contribution to the Raman scattering spectrum of these systems in a magnetic field and the interplay between the valley pseudo-spin and the real spin will be searched for.

Acknowledgement

We would like to thank I. Breslavetz for his help with experiments, and D.M. Basko, S. Berciaud and P. Kossacki for useful discussions. Part of this work has been supported by the graphene flagship project (604391), by the European Research Council (ERC-2012-AdG-320590-MOMB), and by the NCN-Poland grant No. DEC-2013/10/M/ST3/00791.

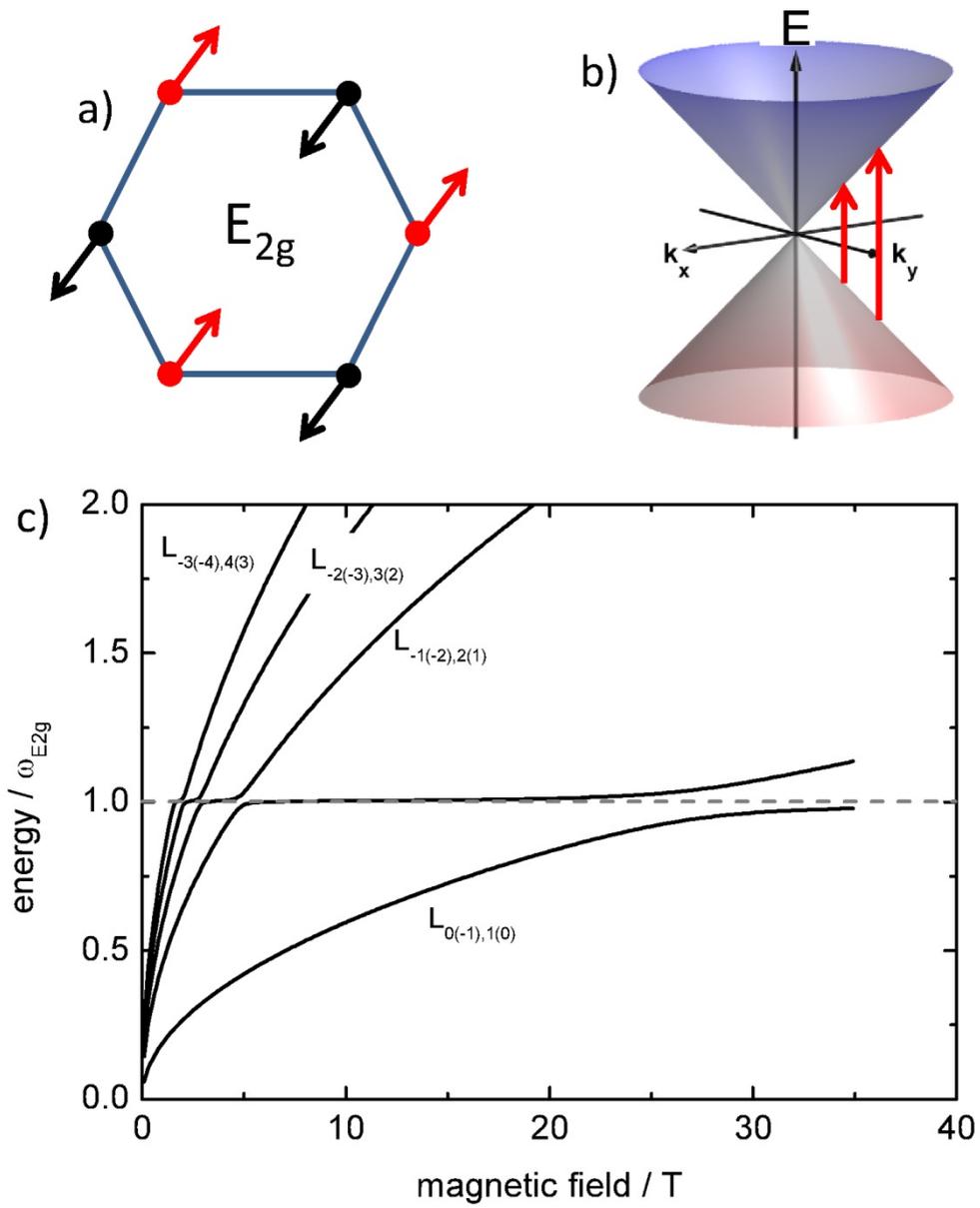

Fig 1: a) $E_{2g}$ phonon displacement pattern, b) Graphene's conical band structure close to K and K' points, Coupled inter Landau level optical phonon modes as a function of the magnetic field as predicted in Ref. [22].

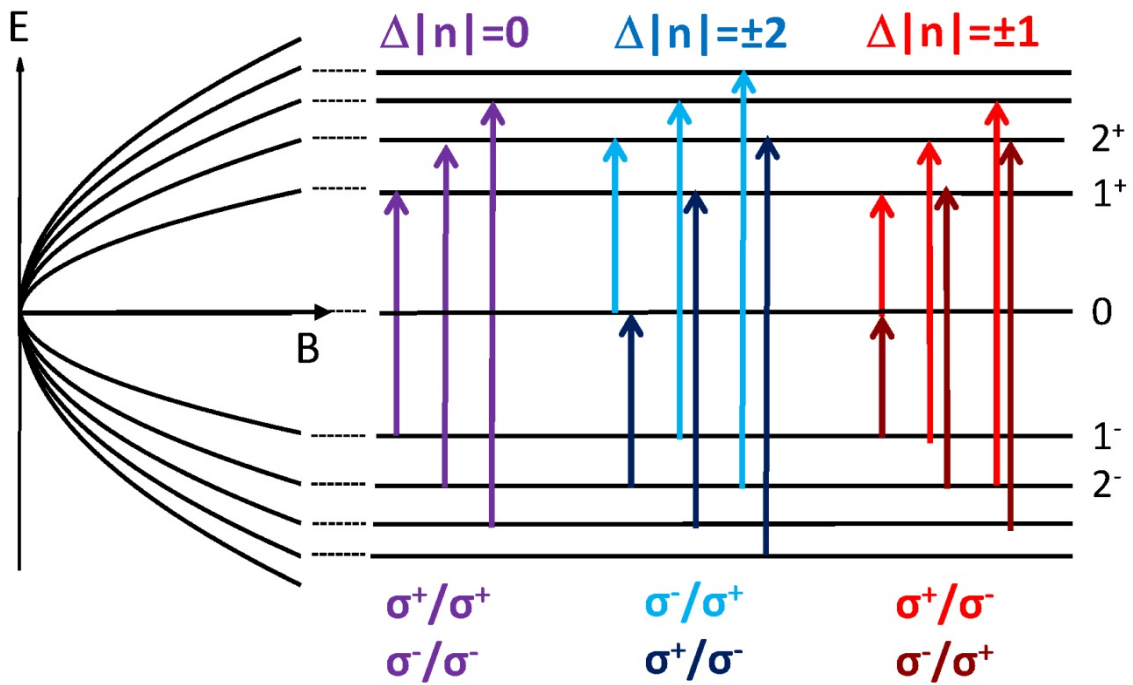

Fig 2: Evolution of Landau levels in graphene as a function of the magnetic field and Raman scattering selection rules. The change of Landau level indices are indicated, together with the corresponding circular polarization configuration σ±/σ±.

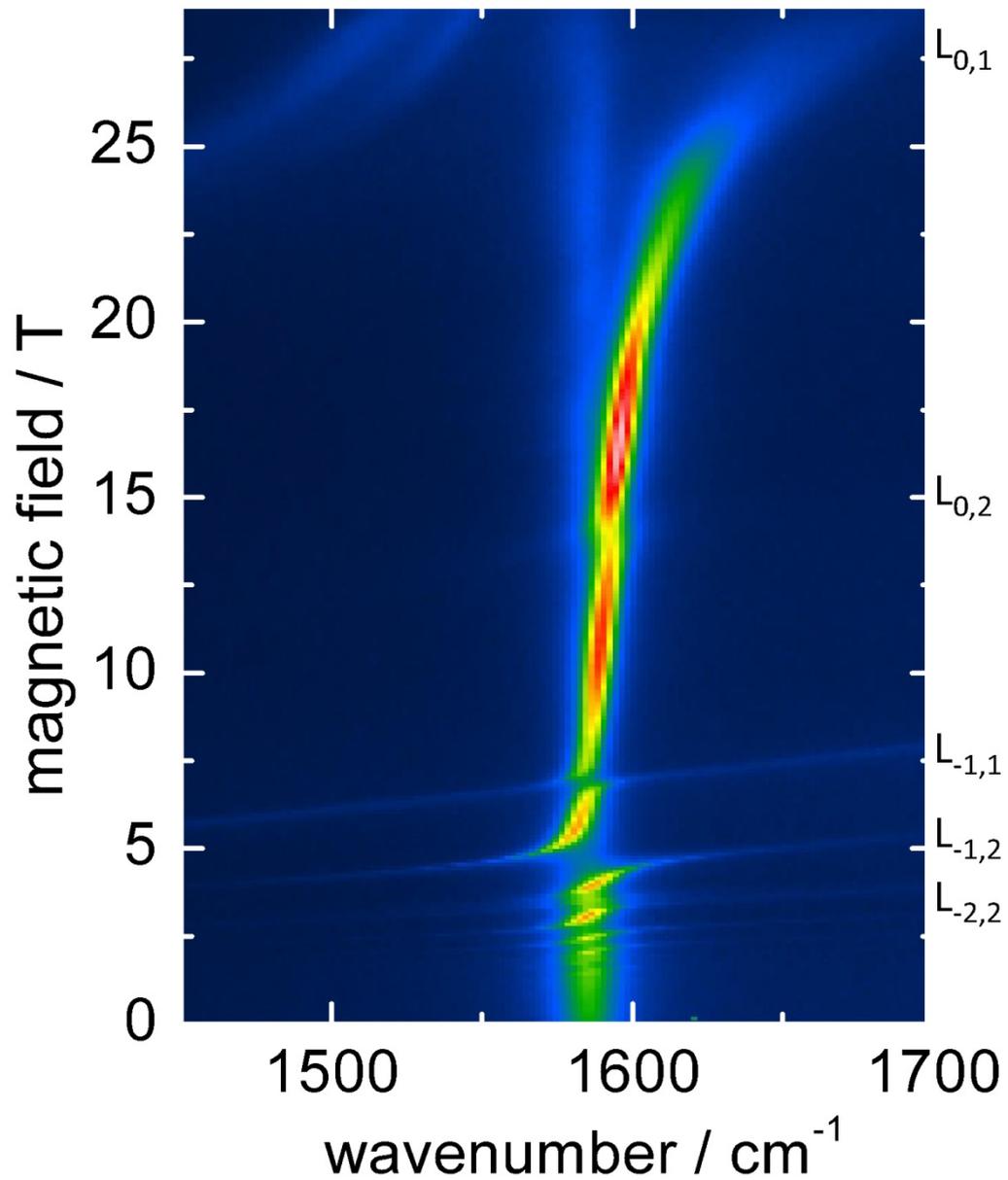

Fig3: False color map of the low temperature scattered intensity in the range of the G band of decoupled graphene flakes on the surface of bulk graphite, as a function of the magnetic field. The different electronic excitations involved in the MPR are indicated on the right side of the figure. Same data as in Ref. [39].

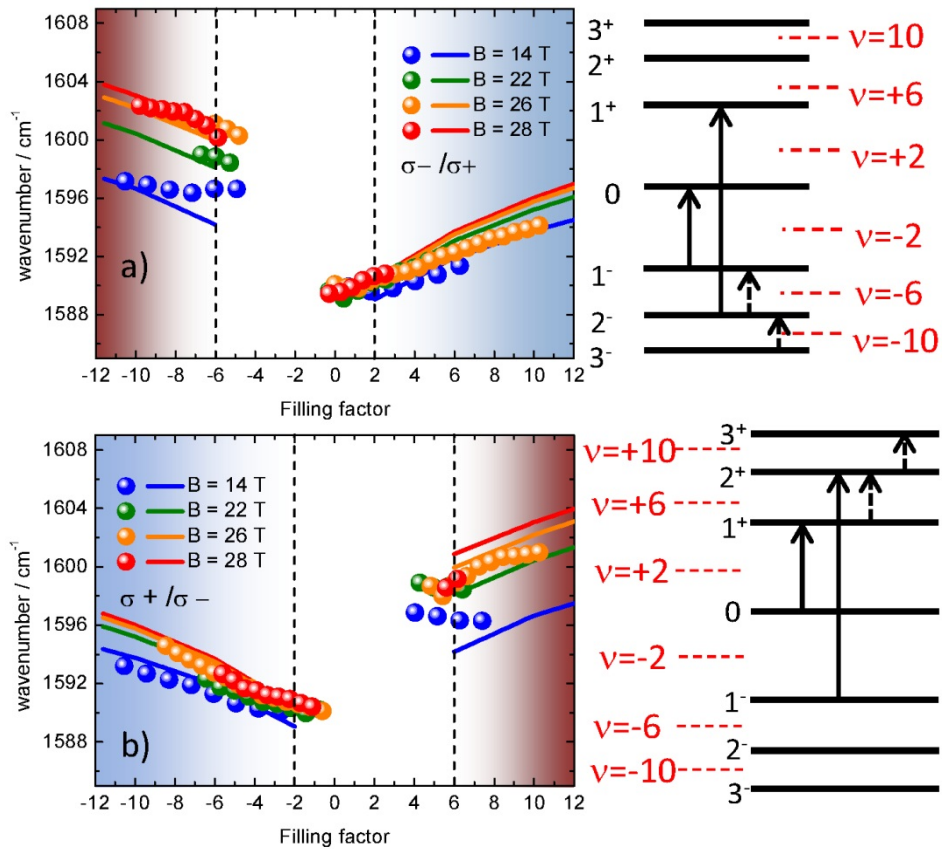

Fig.4: Energy of the $E_{2g}$ phonon of gated graphene as measured in polarization resolved Raman scattering experiment a) and b), as a function of the filling factor, for different values of the magnetic field. The solid lines are calculated phonon energies. The asymmetry between between positive and negative values of the filling factor is representative of the intraband excitation, the cyclotron resonance (dashed arrows in the schematics of possible electronic excitations), to the phonon energy. Reprinted with permission from P. Leszczynski et al., NanoLett. 14, 1460, (2014) [46]. Copyright 2014 American Chemical Society.

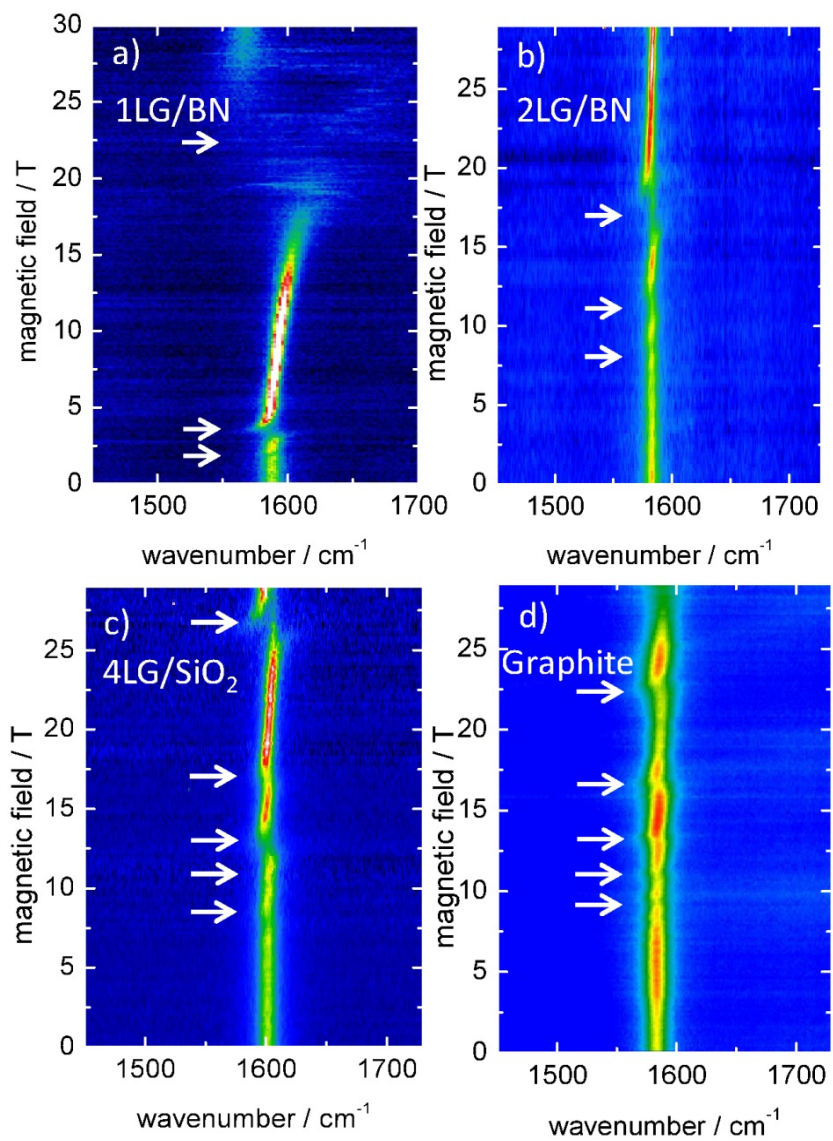

Fig. 5: False color map of the scattered intensity at the G band energy for monolayer graphene encapsulated in hBN, b) bilayer graphene encapsulated in hBN, c) for quadrilayer graphene on $SiO_2$/Si and d) bulk natural graphite. The white arrows indicate the anti-crossings at the resonant magnetic field.

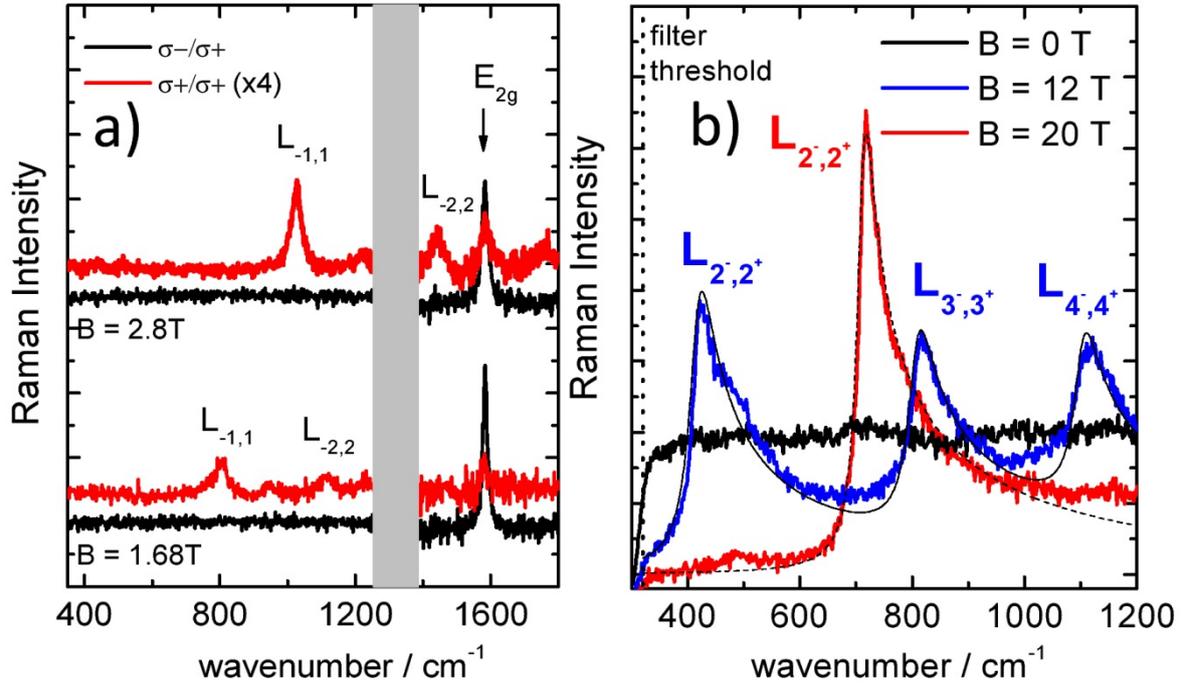

Fig. 6: a) Raman scattering spectra measured at B=2.8T and at B=1.68T on graphene on graphite in the two co and crossed circular polarization configuration showing the Lorentzian shaped inter Landau level excitations. Reprinted with permission from [C. Faugeras, Phys. Rev. Lett.. 107, 06807, (2011)] [37]. Copyright 2011 American Physical Society. b) Raman scattering spectra measured at B=0, 12 and 20T on bulk graphite in the co-circular polarization configuration. The solid and dashed black thin lines are calculations of the joint density of states. Reprinted with permission from [P. Kossacki et al., Phys. Rev. B 84, 235138, (2011)] [29]. Copyright 2011 American Physical Society.

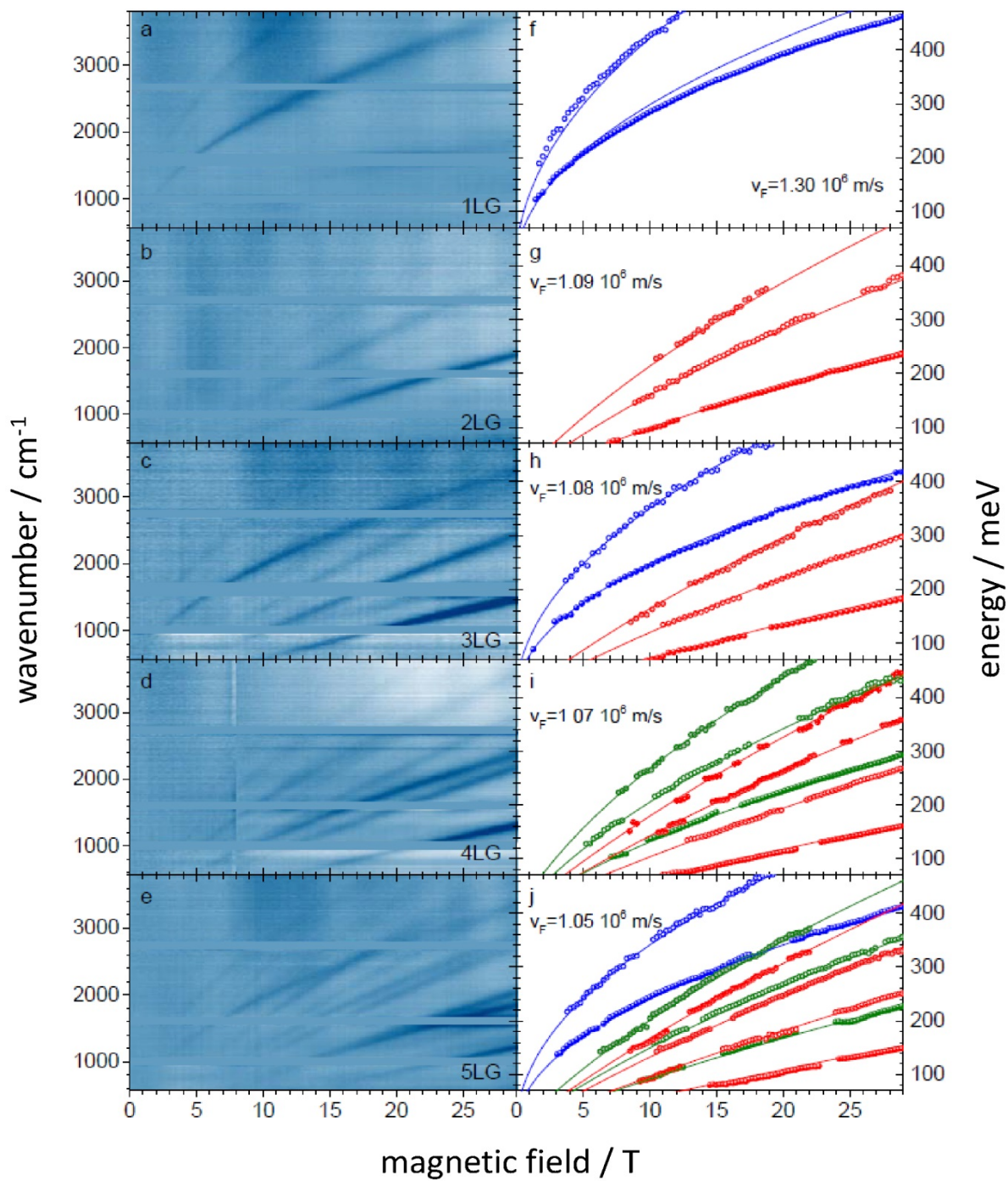

Fig. 7: a-e) False color map of the scattered intensity as a function of the magnetic field for mono- to penta-layer graphene. f-j) Corresponding peak frequencies of the electronic Raman features extracted from Lorentzian fits, together with calculated dispersions. Values for the band velocities are indicated in each panels. Reprinted with permission from S. Berciaud et al., NanoLett. 14, 4548, (2014) [60]. Copyright 2014 American Chemical Society.

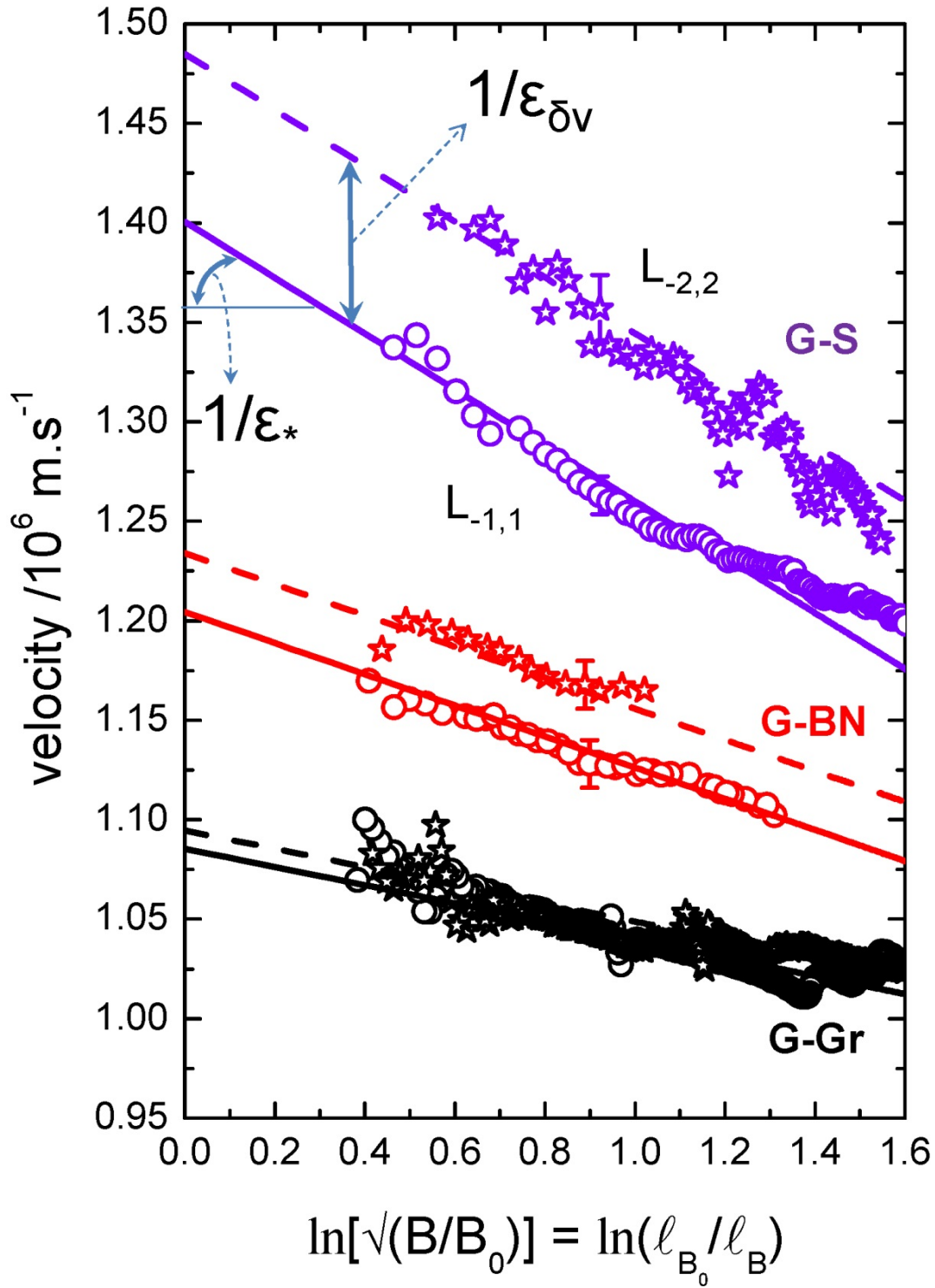

Fig. 8: Evolution of the band velocities extracted for $L_{-1,1}$ and $L_{-2,2}$ for suspended graphene (S-G), graphene encapsulated in hBN (G-BN) and graphene on graphite (G-Gr) as a function of the magnetic field Reprinted with permission from [C. Faugeras et al., Phys. Rev. Lett. 114, 126804, (2015)] [62]. Copyright 2015 American Physical Society.

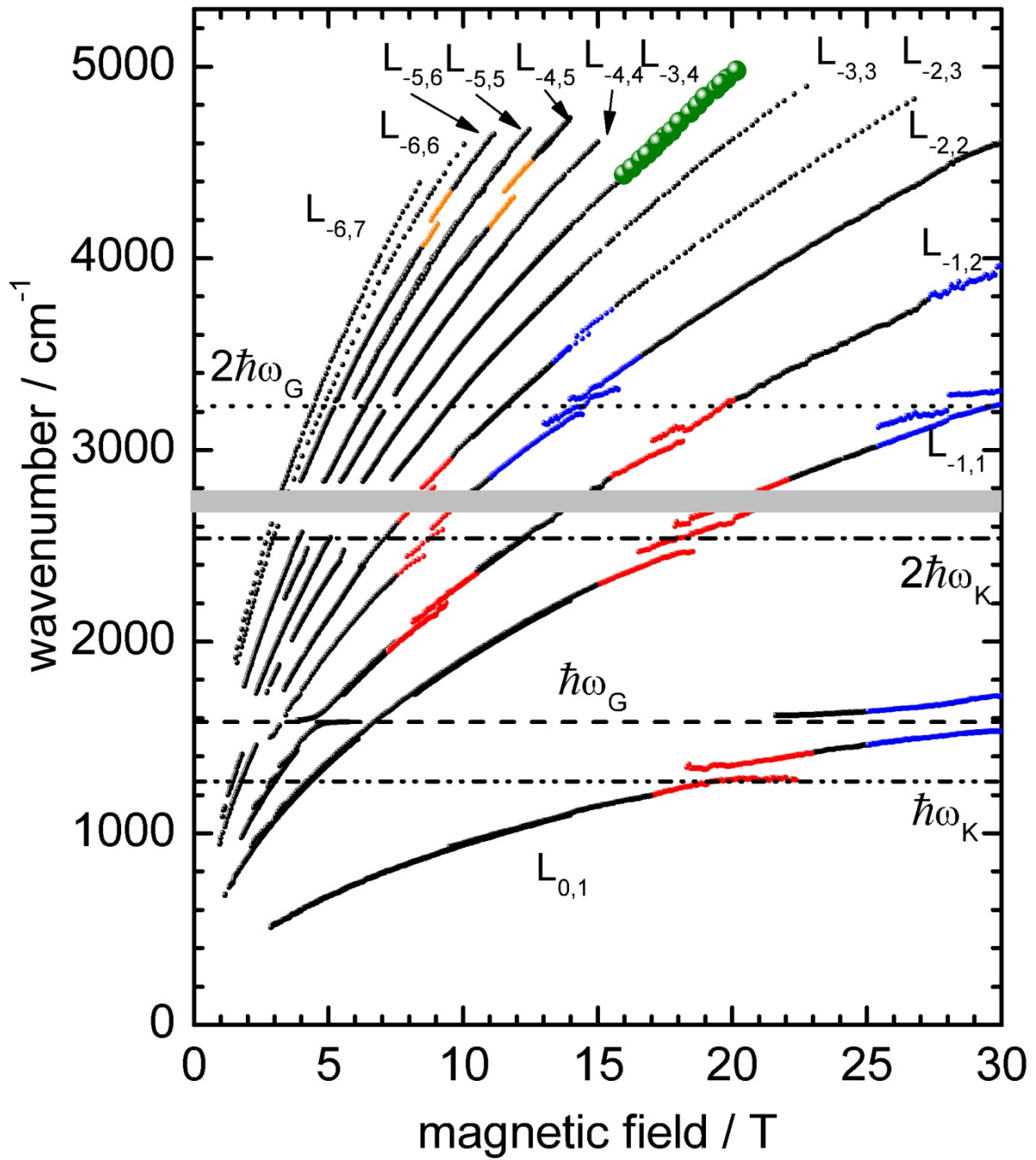

Fig. 9: Magnetic field dependence of the main electronic excitations observed in the magneto-Raman scattering spectra of graphene on graphite. Red points indicate resonances involving K-point phonons, blue point indicate resonances involving Γ-point phonons. Reprinted with permission from [D.M. Basko et al., 2D Materials 3, 015004, (2016)] [89]. Copyright 2016 IOP.